\documentclass[%
 reprint,
 amsmath,amssymb,
 aps,
prl,
]{revtex4-2}
	\usepackage{feynmf}
    \usepackage{float}
	\usepackage{graphics}
	\usepackage{epsfig}
	\usepackage{graphicx}
	\usepackage{color}
	\usepackage{comment}
 \usepackage[scr=boondoxo]{mathalpha}
\usepackage{multirow}
\newcommand*{\rom}[1]{\expandafter\romannumeral #1}

\begin{document}
	\unitlength = 1mm

\title{Dynamical Effective Hamiltonian Approach to Second-Harmonic Generation in Quantum Magnets: Application to NiI$_2$}

\author{Banasree S. Mou}
\affiliation{Department of Physics and Center for Functional Materials, Wake Forest University, Winston-Salem, North Carolina 27109, USA}

\author{Stephen M. Winter}
\email{winters@wfu.edu}
\affiliation{Department of Physics and Center for Functional Materials, Wake Forest University, Winston-Salem, North Carolina 27109, USA}
\date{\today}

\begin{abstract}
Although second harmonic generation (SHG) is a promising and widely used method recently for studying 2D magnetic materials, the quantitative analysis of the full SHG tensor is currently challenging. In this letter, we describe a first-principles-based approach towards quantitative analysis of SHG in insulating magnets through formulation in terms of dynamical effective operators. These operators are  computed by solving local many-body cluster models. We benchmark this method on NiI$_2$, a multiferroic 2D van der Waals antiferromagnet, demonstrating quantitative analysis of reported Rotational Anisotropy (RA)-SHG data. SHG is demonstrated to probe local ring-current susceptibilities, which provide sensitivity to short-range chiral spin-spin correlations. The described methods may be easily extended to other non-linear optical responses and materials. 
\end{abstract}

\maketitle

\noindent {\it Introduction:} Nonlinear spectroscopy comprises a range of experimental probes of higher order light-matter coupling processes, which offer unique insight into ground and excited states of materials \cite{mukamel1999principles,armstrong1962interactions,orenstein2021topology, wan2000nonlinear}. 
A prominent example is Second Harmonic Generation (SHG), in which a sample is illuminated with light of frequency $\omega$, and outgoing light with frequency 2$\omega$ is measured as a function of polarization of incident and outgoing beams. Due to its ability to detect spatially-resolved symmetry and spectral changes \cite{fiebig2005second,ma2020rich,xiao2023classification,wu2024magneto,huang2024second}, as well as applicability to small samples, SHG has emerged as a valuable non-contact probe of bulk and 2D magnets \cite{fiebig1994second,frohlich1998nonlinear,lottermoser2002symmetry,sun2019giant,chu2020linear,shan2021giant,wang2023exciton,lee2021magnetic,liu2022three,guo2023extraordinary,hou2024extraordinary}, superconductors \cite{zhao2017global,jung2024spontaneous}, and materials hosting exotic orders \cite{ZHAO2018207,zhao2016evidence,lei2018observation,petersen2006nonlinear,harter2017parity}.

\begin{figure}[t]
  \includegraphics[width=0.99\linewidth]{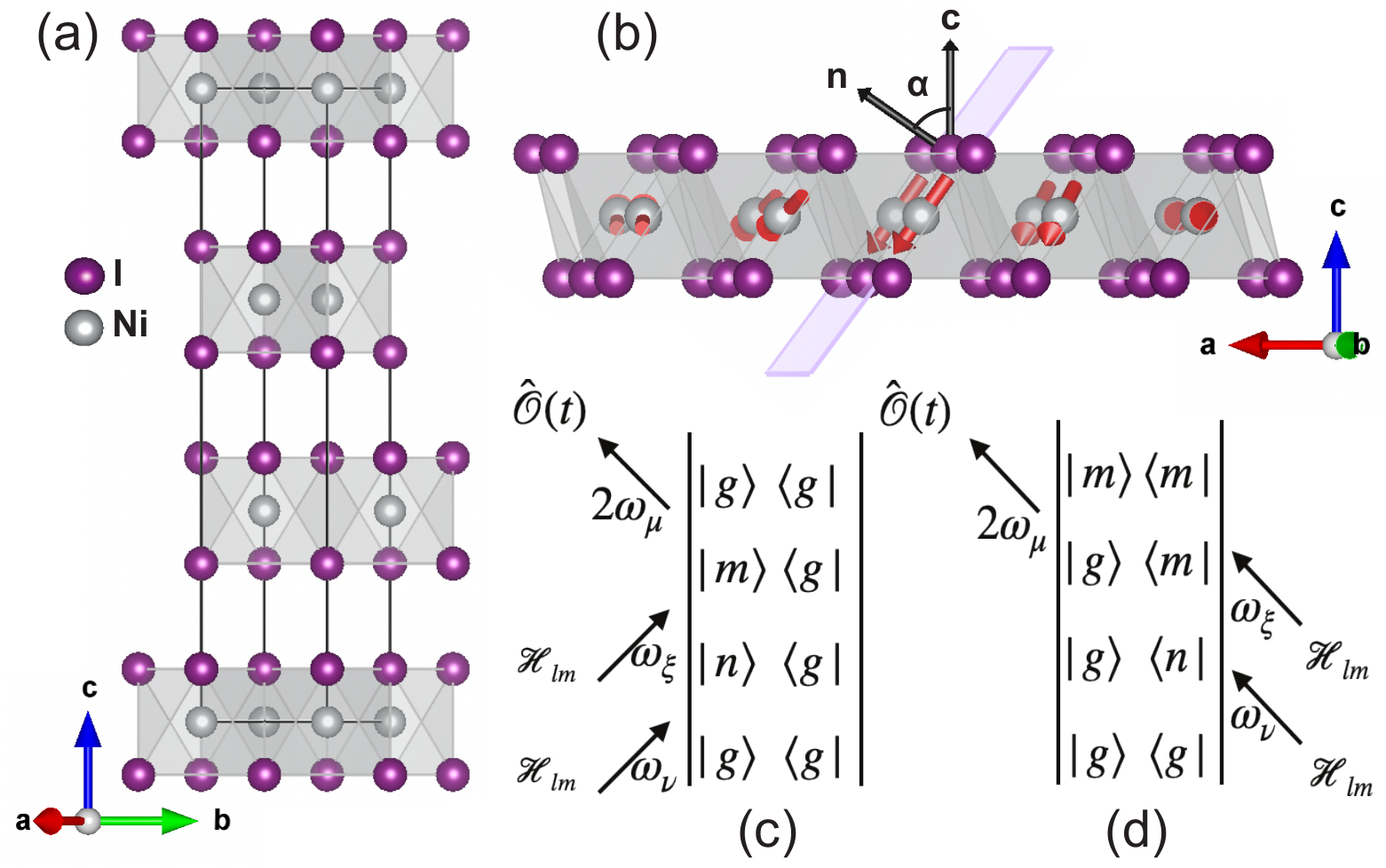}\\
  \caption{(a) Unit cell of NiI$_2$ showing Van der Waals stacked structure. (b) Helical magnetic structure showing spiral plane tilted by angle $\alpha$ from $c$-axis. (c,d) Double-sided Feynman diagrams describing SHG process.}
  \label{fig:structure}
\end{figure}

 Of these materials, NiI$_2$ represents an intriguing case. It crystallizes in the centrosymmetric $\bar{R}3m$ space group, featuring triangularly arranged edge-sharing Ni$^{2+}$ octahedra ($3d^{8}$, S = 1). In bulk, NiI$_2$ undergoes two successive magnetic transitions, settling into a helimagnetic state below $T_{N2} = 58$ K \cite{billerey1977neutron,kuindersma1981magnetic,ju2021possible, son2022multiferroic, liu2024competing}. In this phase, spins rotate in a plane canted by an angle $\alpha = 55^\circ \pm 10^\circ$ from the crystallographic $c$-axis, with an incommensurate in-plane wavevector $\mathbf{Q}_1 = (0.138,0)$, as depicted in Fig.~\ref{fig:structure}(b). The helimagnetic order spontaneously breaks parity symmetry, which leads to both a large SHG response \cite{ju2021possible,song2022evidence,wu2023layer}, and a finite electrical polarization \cite{kurumaji2013magnetoelectric} (type-II multiferroicity). While the magnetic symmetry can be deduced from the polarization-dependent SHG response, the microscopic mechanisms contributing to SHG remain debated \cite{jiang2023dilemma,song2023reply}. These discussions highlight that most SHG studies of quantum materials employ \emph{qualitative} analysis to identify e.g.~order parameter symmetries, and thus do not utilize the full \emph{quantitative} information in the measured non-linear susceptibility tensors. This has motivated calls for  quantitative theoretical methods to predict non-linear responses in quantum materials, and link them to microscopic properties \cite{orenstein2021topology,ZHAO2018207}.

As discussed in this letter, a major challenge for modeling non-linear optical responses of magnetic insulators is the accurate treatment of many-body correlations within both the low-energy (spin) states and high energy electronic excited states. Motivated by previous microscopic analyses of SHG \cite{muthukumar1995microscopic,muthukumar1996theory,muto1998magnetoelectric,iizuka2001second}, we address this by introducing dynamical effective operators representing the low-energy effects of optical coupling to electronic excitations. We then describe a first-principles-based approach to calculate these operators using exact diagonalization of local many-body cluster models, which has proved valuable for computing magnetic \cite{winter2016challenges,riedl2019ab,riedl2022microscopic,xiang2023disorder}, magneto-elastic \cite{dhakal2024spin}, and exciton-magnon couplings \cite{dhakal2024hybrid}. We apply this approach to NiI$_2$, and identify the excited state processes contributing to the computed SHG operators. Finally, using these operators, we demonstrate accurate extraction of the spin orientation in NiI$_2$ from experimental SHG data, which represents a quantity that cannot be deduced solely from symmetry analysis. This validates the described approach and demonstrates potential as a general tool for \emph{quantitative} analysis of non-linear optical responses in magnetic insulators and other correlated materials.

\noindent {\it SHG Operators:} Consider a material with a Hamiltonian $\mathcal{H}_0$ and light-matter coupling $\mathcal{H}_{\rm lm}$ in the electric-dipole approximation: 
\begin{align}
    \mathcal{H}_{\rm lm} = \sum_\nu\int d\omega \ \hat{\vec{\mathcal{L}}}(\omega)\cdot \vec{E}(\omega) \ e^{i\omega t}
\end{align}
where $\vec{E}(\omega)$ is the electric field of the light of frequency $\omega$ and $\mathcal{L}(\omega)$ are dipole/current coupling operators (see end matter). In SHG, two incoming photons of frequency $\omega$ and polarizations $\nu$ and $\xi$ are absorbed, and a photon is emitted with frequency $2\omega$ and polarization $\mu$. The electric field of the outgoing radiation is given by $E_{\rm out}^\mu(2\omega)\propto \sum_{\xi,\nu}\int d\omega \ \chi_{\mu\xi\nu}(2\omega; \omega,\omega) E_{\rm in}^\xi (\omega) E_{\rm in}^\nu (\omega)$, with the SHG susceptibility:
\begin{align}
 \chi_{\mu\xi\nu} (2\omega;\omega,\omega)    \propto i\omega \left\langle \hat{\mathcal{O}}_{\mu\xi\nu}(\omega)+\hat{\mathcal{O}}_{\mu\xi\nu}^\dagger(-\omega)\right\rangle_\beta\label{eq:chi2tr}
\end{align}
where $\langle ... \rangle_\beta = \frac{1}{Z}\text{Tr}[e^{-\beta \mathcal{H}} ... ]$. A key observation is that $\chi_{\mu\xi\nu}$ is the thermodynamic expectation value of material-dependent SHG operators $\hat{\mathcal{O}}_{\mu\xi\nu}$, which may be determined separately from the evaluation of $\chi_{\mu\xi\nu}$. For magnetic insulators, it is appropriate to write the SHG operators in the low-energy spin-space explicitly in terms of spin operators. Similar approaches are commonly applied to analyze magnetic Raman scattering\cite{fleury1968scattering,yang2021non, mou2024comparative}. In principle, this may be accomplished by explicitly computing the two contributions depicted in the double sided Feynman diagrams in Fig.~\ref{fig:structure}(c,d): 
\begin{align}
\hat{\mathcal{O}}_{\mu\xi\nu}(\omega) =  \hat{\mathbb{P}}\sum_{n,m}
\frac{\hat{\mathcal{L}}^\nu (\omega)|n\rangle\langle n|}{\hbar\omega+\mathcal{H}_0-E_n-i\eta} \cdot
\hspace{20mm}\nonumber \\
\cdot\left( \frac{\hat{\mathcal{L}}^\xi(\omega)|m\rangle  \langle m|\hat{\mathcal{L}}^\mu(-2\omega)}{2\hbar\omega+\mathcal{H}_0-E_m-i\eta}
- \frac{\hat{\mathcal{L}}^\mu(-2\omega)|m\rangle  \langle m|\hat{\mathcal{L}}^\xi(\omega)}{2\hbar\omega+E_m-E_n-i\eta}
 \right)\hat{\mathbb{P}} \label{eq:O}
 \end{align}
 where $\hat{\mathbb{P}}$ is a projection operator onto the low-energy space and  $|n\rangle$ is an eigenstate of $\mathcal{H}_0$ with energy $E_n$. For optical frequencies, the relevant excited states $|n\rangle,|m\rangle$ include both intersite $d$-$d$ and ligand-metal charge transfer (LMCT) excitations, depicted in Fig.~\ref{fig2}(m).  
 
 For pure electric-dipole processes, the SHG operators $\hat{\mathcal{O}}_{\mu\xi\nu}$ transform as electric octupoles (parity odd, time-reversal even). Local crystalline symmetries place constraints on their form. In general, they may be expanded in terms of spin dipole operators as:
\begin{align}
\hat{\mathcal{O}}_{\mu\xi\nu}(\omega) =  
     f^{\mu\xi\nu}(\omega) + \sum_i \mathbf{G}_i^{\mu\xi\nu}(\omega) \cdot \mathbf{S}_i
\nonumber \\
 + \sum_{ij} \mathbf{S}_i \cdot \mathbb{H}_{ij}^{\mu\xi\nu}(\omega) \cdot \mathbf{S}_j
    + ... \label{eq:Ospin}
\end{align}
where $\mu,\xi,\nu\in\{x,y,z\}$ and $f$, $\mathbf{G}$ and $\mathbb{H}$ are frequency- and polarization-dependent parameters. For NiI$_2$, the centrosymmetric $\bar{R}3m$ structure forbids spin-independent contributions ($f^{\mu\xi\nu} = 0$). Similarly, there is no linear dependence of SHG on a magnetic order parameter ($\mathbf{G}^{\mu\xi\nu} = 0$), because the magnetic Ni ions lie on inversion centers, making all such spin operators parity-even and therefore forbidden. In the absence of structural distortions, only select $\mathbb{H}_{ij}^{\mu\xi\nu}$ terms that probe short ranged chiral spin-spin correlations are allowed by symmetry as discussed below.

\begin{figure}[b]
  \includegraphics[width=0.9\linewidth]{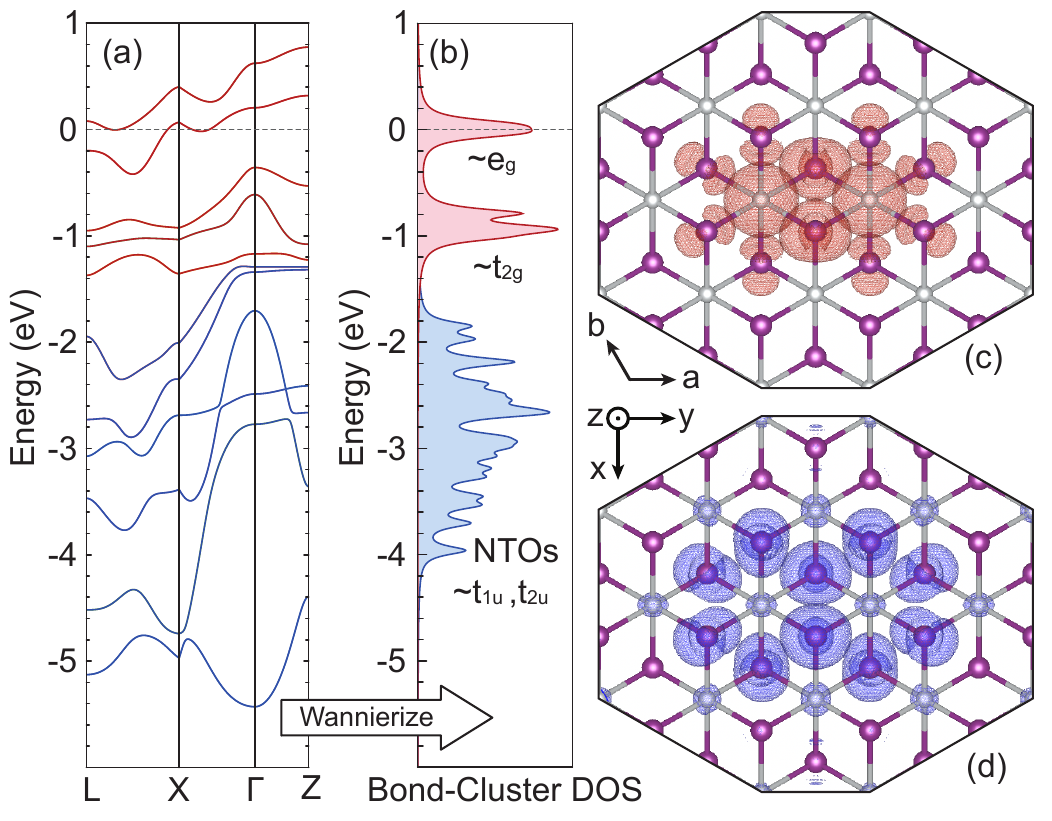}\\
  \caption{(a) DFT bandstructure of NiI$_2$. (b) One-particle density of states for nearest neighbor cluster model including Ni $d$-orbital Wannier functions and ligand Natural Transition Orbitals (NTOs). (c,d) Total density $\sum |\psi(\vec{r})|^2$ of (c) $d$-orbital WFs and (d) NTOs.
  }
  \label{fig:orb}
\end{figure}

\begin{figure*}[t]
  \includegraphics[width=\linewidth]{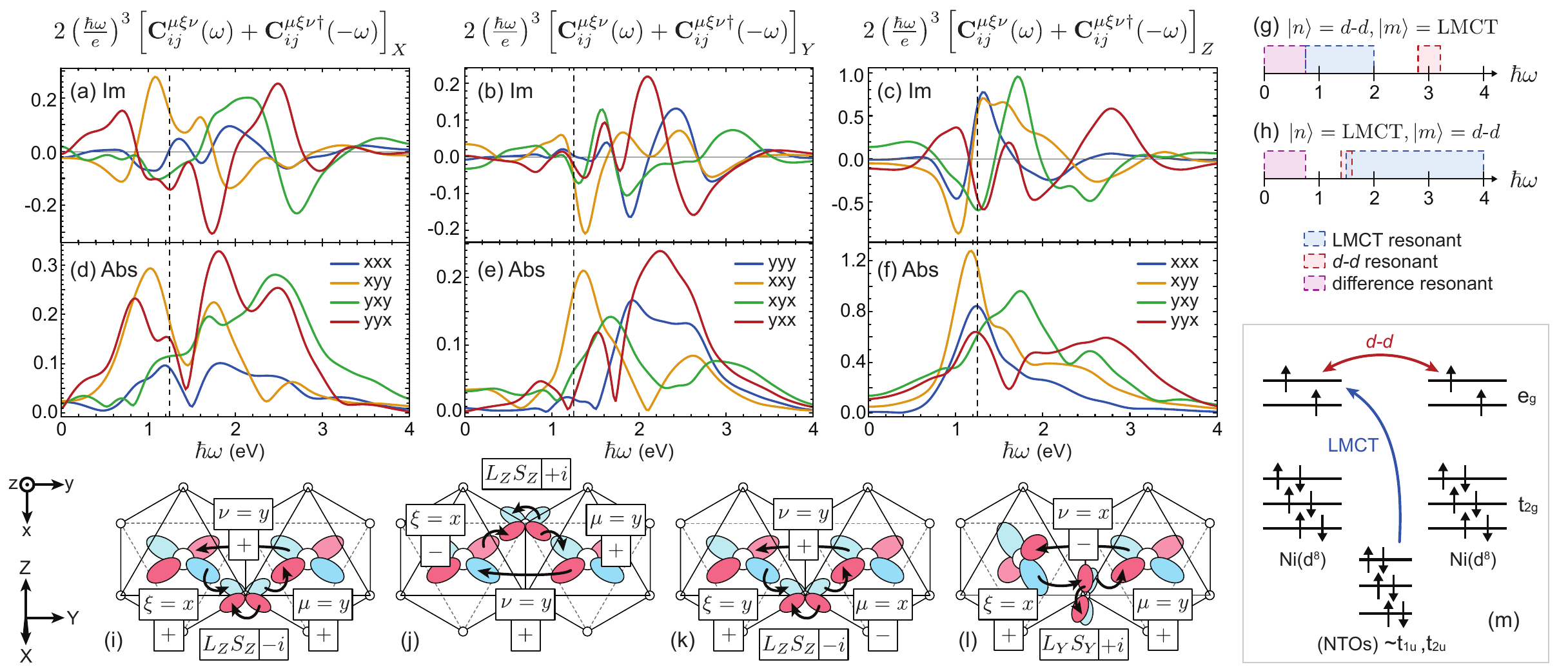}\\
  \caption{(a-f) Computed ($X,Y,Z$) components of the  SHG operator couplings presented as (a-c) 
$2(\hbar\omega/e)^3\text{Im}[\mathbf{C}_{ij}^{\mu\xi\nu}(\omega)+\mathbf{C}_{ij}^{\mu\xi\nu\dagger}(-\omega)]$
  and (d-f) $2(\hbar\omega/e)^3\text{Abs}[\mathbf{C}_{ij}^{\mu\xi\nu}(\omega)+\mathbf{C}_{ij}^{\mu\xi\nu\dagger}(-\omega)]$ in units of \AA$^3$. Polarization indices are given in ($x,y,z$) coordinates. (g,h) Energy ranges for resonant SHG, referring to eq'n (\ref{eq:O}). (i-l) Example ring current excitation pathways contributing to SHG operators at $\hbar\omega = 1.25$ eV. Arrows refer to the transfer of holes. The relative signs of $\langle g|\hat{\mathcal{L}}^\nu(\omega)|n\rangle$, $\langle n|\hat{\mathcal{L}}^\xi(\omega)|m\rangle$, and $\langle m|\hat{\mathcal{L}}^\mu(-2\omega)|g\rangle$ matrix elements are indicated. (m) Depiction of LMCT and intersite $d$-$d$ excitations. \label{fig2} 
    }
\end{figure*}

\noindent {\it Computational Approach:} We propose a general approach for computing $\hat{\mathcal{O}}_{\mu\xi\nu}$ by numerically evaluating eq'n (\ref{eq:O}) for many-body cluster models including the sites of interest. To incorporate LMCT contributions, we include ligand orbitals using a minimal set of Natural Transition Orbitals \cite{martin2003natural} (NTOs). Full implementation details employing Fleur \cite{fleurCode,fleurWeb} and wannier90  \cite{freimuth2008maximally,mostofi2014updated} are given in the End Matter; here we sketch relevant details. 

For a given local cluster, we first obtain one-particle contributions to the material Hamiltonian $\mathcal{H}_0$. Wannier functions (WFs) are first constructed utilizing $d$-orbital projections to fit the bands near the Fermi energy; the resulting WFs are antibonding combinations of $d$- and $p$-orbitals hereafter denoted as metal-centered orbitals [their complement within the full $(d+p)$ basis are denoted ligand-centered orbitals]. Transition matrix elements of $\hat{\mathcal{L}}^x$, $\hat{\mathcal{L}}^y$, and/or $\hat{\mathcal{L}}^z$ are then computed between the metal-centered orbitals within the cluster and ligand-centered orbitals in the entire supercell utilized in the WF calculation. Performing a singular value decomposition on the transition matrices and orthonormalizing the resulting vectors yields a basis of ligand-centered NTOs, with at most three times as many NTOs as cluster $d$-orbitals. These NTOs span the entire optical spectral weight of the LMCTs.

To obtain $\hat{\mathcal{O}}_{\mu\xi\nu}(\omega)$ for a given cluster, we supplement the single-particle terms in $\mathcal{H}_0$ with on-site Coulomb interactions on the metals. A double-counting correction is applied in the AMF scheme \cite{czyzyk1994local,petukhov2003correlated}. 
We diagonalize the cluster Hamiltonian in the truncated $d+$NTO many-body Fock space, and project the low-energy eigenstates onto pure spin states to establish a mapping to spin operators. Finally, we compute eq'n (\ref{eq:O}) in the electronic Fock basis  for each polarization combination  using standard Krylov space methods\cite{meyer1989band,weikert1996block}, and apply the spin operator mapping to obtain $\hat{\mathcal{O}}_{\mu\xi\nu}$.

For NiI$_2$, we use in-plane polarized $\hat{\mathcal{L}}^x$, $\hat{\mathcal{L}}^y$ to construct NTOs to capture relevant orbitals for modeling normal incidence SHG [see Fig.~\ref{fig:orb} for ($x,y,z$) coordinate definition]. Fig.~\ref{fig:orb}(a,b) compares the computed band structure with the one-electron DOS in the truncated $d$+NTO basis for a nearest neighbor bond cluster, demonstrating adequate correspondence.  Fig.~\ref{fig:orb}(c,d) shows the total density $\sum|\psi(\vec{r})|^2$ of the metal-centered WFs and ligand-centered NTOs. The metal $d$-orbital WFs are split by the octahedral crystal field into nominal $e_g$ and $t_{2g}$ combinations with significant Ni and I character. The NTOs represent combinations of I $p$-orbitals with nominal $t_{1u}$ and $t_{2u}$ symmetry and therefore little density on the Ni. Each two-site $d^8$ Ni cluster includes 20 metal spin-orbitals and 40 spin-NTOs, occupied by 56 electrons (487,635 Fock states in total), which is easily computationally tractable for Krylov space methods.

\noindent {\it NiI$_2$ Results:} Due to the small magnitude of further neighbor optical matrix elements, we find the SHG operators are dominated by nearest neighbor terms. The $C_{2h}$ point group symmetry of each bond constrains these to take the form: 
\begin{align}
    \hat{\mathcal{O}}_{\mu\xi\nu}(\omega) = \sum_{\langle ij\rangle}\mathbf{C}_{ij}^{\mu\xi\nu} (\omega) \cdot (\mathbf{S}_i\times \mathbf{S}_j)
\end{align}
where $\mathbf{C}_{ij}^{\mu\xi\nu} (\omega)$ is a complex polar vector that is a dynamical analogue of the Dzyalloshinskii-Moriya (DM) vector. It is useful to discuss this vector in the mixed coordinates shown in Fig.~\ref{fig2}. Polarization indices are indicated in the bond-dependent $(x,y,z)$ coordinates with $y$ parallel to the bond, and $z || c$. $\mathbf{C}$-vector components are indicated in the rotated $(X,Y,Z)$ coordinates with $Z$ normal to the edge-sharing plane. By symmetry, $[\mathbf{C}_{ij}^{\mu\xi\nu}]_X$ and $[\mathbf{C}_{ij}^{\mu\xi\nu}]_Z$ are finite for $\mu\xi\nu = xxx$, $xyy$, $yxy$, and $yyx$. $[\mathbf{C}_{ij}^{\mu\xi\nu}]_Y$ is finite for $\mu\xi\nu = yyy$, $yxx$, $xyx$, and $xxy$. The complex phase of the $\mathbf{C}$-vectors determines the relative phase shift of outgoing radiation, and is determined both by the phase of the light-matter matrix elements and the proximity of $\omega$ and $2\omega$ to resonance. 

In Fig.~\ref{fig2}(a-f), we show the computed SHG couplings. 
The discrete poles in eq'n (\ref{eq:O}) were broadened with $\eta = 0.5$ eV, which is roughly twice the average separation between adjacent poles. 
To gain insight into the frequency and polarization dependence, it is useful to consider different excitation pathways contributing to eq'n  (\ref{eq:O}). The $\mathbb{H}$ terms require one of the $|n\rangle,|m\rangle$ states to be an intersite $d$-$d$ excitation to generate a dependence on the spins at two different sites. The largest optical matrix elements occur between metal and ligand centered orbitals, implying an LMCT state should be the other excited state. As a consequence, $\mathbb{H}$ terms arise primarily from induced metal-ligand ring currents including those depicted in Fig.~\ref{fig2}(i-l).

In practice, at least one of the denominators in eq'n (\ref{eq:O}) must be resonant to yield a significant contribution to the SHG couplings $\mathbf{C}_{ij}(\omega)$ or $\mathbf{C}_{ij}^\dagger(-\omega)$. 
This leads to several possibilities: (i) $|n\rangle$ = $d$-$d$, $|m\rangle = $ LMCT and $\pm\hbar\omega \approx \Delta E_{dd}$ or $\pm2\hbar\omega \approx \Delta E_{\rm LMCT}$ or $\pm2\hbar\omega \approx \Delta E_{dd}-\Delta E_{\rm LMCT}$, (ii) $|n\rangle$ = LMCT, $|m\rangle = $ $d$-$d$ and $\pm\hbar\omega \approx \Delta E_{\rm LMCT}$ or $\pm2\hbar\omega \approx \Delta E_{dd}$ or $\pm2\hbar\omega \approx \Delta E_{\rm LMCT}-\Delta E_{dd}$. For our model, intersite $d$-$d$ excitations appear at $\Delta E_{dd}\sim 3$ eV and LMCT excitations appear in the range $\Delta E_{\rm LMCT}\sim 1.5$ - 4 eV.  Fig.~\ref{fig2}(g,h) depicts the energy ranges over which resonance conditions are satisfied.

We now focus on $\hbar\omega \approx 1.25$ eV ($\lambda = 991$ nm), which was employed in the experiment of \cite{song2022evidence}. In this case, there is only one resonant process corresponding to the diagram in Fig.~\ref{fig:structure}(c) with $2\hbar \omega \approx \Delta E_{\rm LMCT}$ and $|n\rangle$ = $d$-$d$, $|m\rangle$ = LMCT. A selection of non-vanishing excitation pathways are depicted in Fig.~\ref{fig2}(i-l). For example, Fig.~\ref{fig2}(i) depicts the transfer of holes in a process contributing to $[\mathbf{C}_{ij}^{yxy}]_Z$ in which a photon ($\nu = y$) first couples (non-resonantly) to an excited $e_g\to e_g$ transition moving a hole from site $j$ to $i$. Absorption of a second photon ($\xi=x$) transfers the hole from site $i$ to a ligand yielding an LMCT state. Spin-orbit coupling (SOC) on the ligand shifts the hole to a different $p$-orbital via $L_ZS_Z$. Finally, emission of a $2\omega$ photon ($\mu = y$) transfers the hole back to site $j$. The component of the $\mathbf{C}$ vector to which a pathway contributes is determined by the SOC component acted at the ligand. The relative signs of the light-matter and spin-orbit matrix elements are indicated. Fig.~\ref{fig2}(j) shows another pathway contributing to $[\mathbf{C}_{ij}^{yxy}]_Z$ involving the second edge-sharing ligand, demonstrating the product of matrix elements has the same sign, and therefore adds constructively. Fig.~\ref{fig2}(k) shows a contribution to $[\mathbf{C}_{ij}^{xyy}]_Z$ that has opposite sign to $[\mathbf{C}_{ij}^{yxy}]_Z$, consistent with the computed values near $\hbar\omega \approx 1.25$ eV.
 
Finally, Fig.~\ref{fig2}(l) depicts a process contributing to $[\mathbf{C}_{ij}^{yxx}]_Y$ involving an initial $t_{2g}\to e_g$ transition. Similar processes contribute to both $\mathbf{C}_X$ and $\mathbf{C}_Y$ but are suppressed by smaller $\mathcal{L}$ matrix elements between $e_g$ and $t_{2g}$ orbitals, thus explaining the larger magnitude of $\mathbf{C}_Z$. These microscopic considerations are consistent with the relative signs and magnitudes of the computed SHG operators (which include all excitation pathways).

\begin{figure}[b]
  \includegraphics[width=\linewidth]{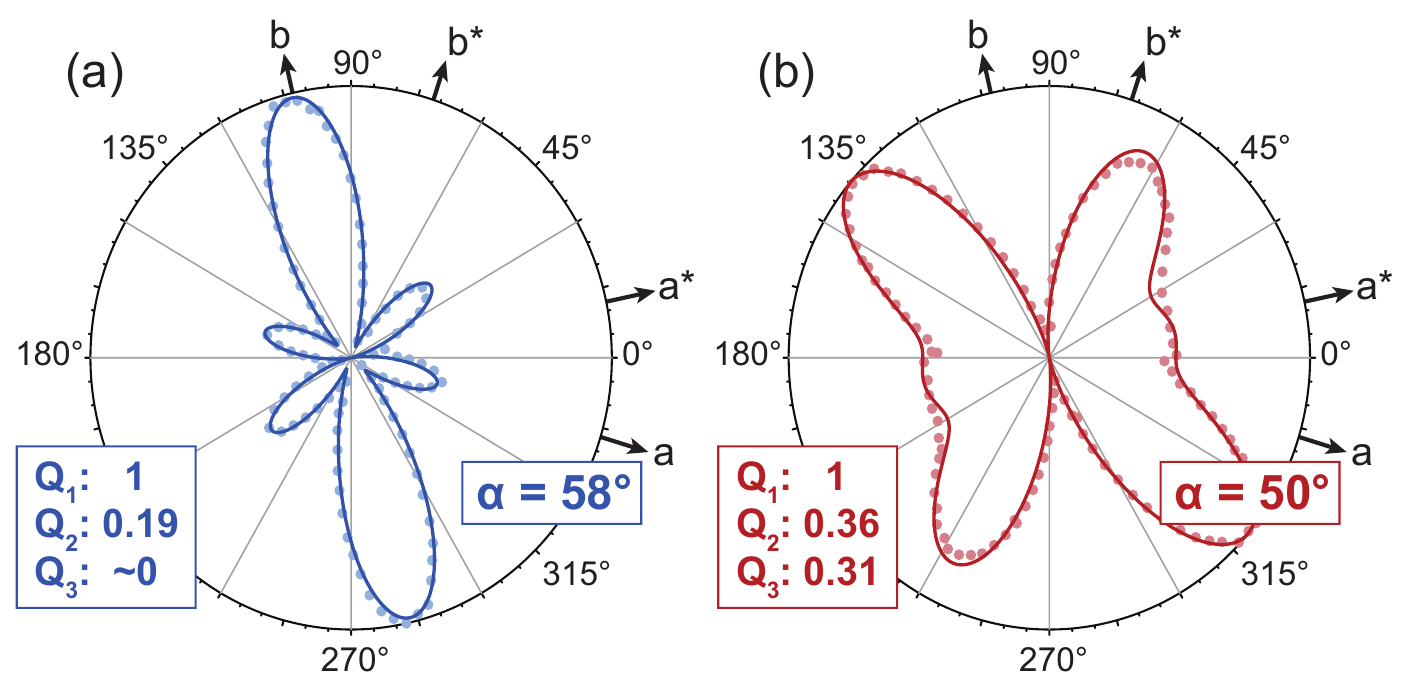}\\
  \caption{Fitted RA-SHG intensity (lines) with respect to polarization angle $\theta$ for best fit spin spiral angle $\alpha$ for (a) parallel, (b) perpendicular polarization. Dots are experimental data on CVD-grown NiI$_2$ crystals with $\omega \approx 1.25$ eV extracted from \cite{song2022evidence}. Fitted $\alpha$ values and relative population of domains with in-plane $\mathbf{Q}_1 = (q,0), \mathbf{Q}_2 = (q,-q), \mathbf{Q}_3=(-q,q)$ are indicated.}
  \label{fig3}
\end{figure}

Lastly, we address the rotational anisotropy SHG (RA-SHG) experiments reported in \cite{song2022evidence}. Using the computed SHG operators, we fit extracted experimental data to $I_{2\omega} \propto |\chi(2\omega;\omega,\omega)|^2$ allowing for mixture of magnetic domains and a variable tilt angle $\alpha$ [defined in Fig.~\ref{fig:structure}(b)]. Fitting results are shown in Fig.~\ref{fig3} for both parallel ($\mu$=$\xi$=$\nu$) and perpendicular ($\mu$$\perp$$\xi$=$\nu$) polarization. The RA-SHG patterns are sensitive to $\alpha$ because the orientation of $\hat{n} \ || \  \langle\mathbf{S}_i \times \mathbf{S}_j\rangle$ selects which components of $\mathbf{C}_{ij}$ contribute to the SHG susceptibility. Best fits are obtained for $\alpha = 58^\circ$ and $\alpha = 50^\circ$, which is in remarkable agreement with the neutron scattering estimate of $\alpha = 55^\circ \pm 10^\circ$ \cite{kuindersma1981magnetic}. This implies $\hat{n}$ is roughly perpendicular to the $Z$-axis of all nearest neighbor bonds with finite $\langle\mathbf{S}_i \times \mathbf{S}_j\rangle$; indeed, good fits are obtained only for a narrow range of $\alpha$ where contributions from $[\mathbf{C}_{ij}^{\mu\xi\nu}]_Z$ are suppressed. This orientation is promoted by large bond-dependent ``Kitaev'' couplings \cite{li2023realistic} anticipated for NiI$_2$ \cite{stavropoulos2019microscopic}. These findings underscore that \emph{quantitative} analysis of RA-SHG using \emph{ab-initio} approaches can extract details beyond the symmetries of order parameters.

\noindent {\it Discussion:} We conclude, in edge-sharing materials, SHG probes induced metal-ligand ring currents, as anticipated in e.g.~\cite{zhao2016evidence}. In NiI$_2$, the associated susceptibility measures short-ranged $\langle \mathbf{S}_i\times \mathbf{S}_j\rangle$ chiral correlations, particularly due to strong SOC of the iodine ligands. For a given material, the precise correlations probed by SHG at a given frequency can be identified \emph{a-priori} using the first principles-based method described in this letter.
This method complements pure-DFT approaches \cite{wang2017giant,sharma2004second,song2020nonreciprocal,li2022probing,wu2024giant} based on summations over Kohn-Sham bands, by (i) incorporating local many-body effects explicitly, and (ii) identifying the low-energy operators rather than directly computing $\chi_{\mu\xi\nu}$. The latter capability aids qualitative interpretation of experimental data, and also facilitates analysis of complex orders by removing the necessity to 
treat those orders at the DFT level.

It may be emphasized that the computed SHG operators do not consider lattice distortions or orbital polarization, which provides insight into the competing interpretations of the NiI$_2$ SHG in \cite{jiang2023dilemma} and \cite{song2023reply}. NiI$_2$ is demonstrated to be multiferroic, implying a weak orbital/lattice parity-breaking accompanies helimagnetic order. Consistently, Ref.~\cite{song2023reply} invoked a spin-independent SHG contribution $f^{\mu\xi\nu} \propto |\mathbf{P}|$ associated with an orbital/lattice polarization $\mathbf{P} \propto \langle \mathbf{S}_i \times \mathbf{S}_j\rangle$, induced by magnetoelectric coupling. While this mechanism leads to similar conclusions regarding dependence of SHG on $\langle \mathbf{S}_i \times \mathbf{S}_j\rangle$, it is not included in our calculations. Instead, magnetic order alone is sufficient to restrict the excitation pathways accessible from the ground state, which produces finite SHG as soon as the low-energy (spin) wavefunction breaks parity. This is the physical origin of the $\mathbf{S}_i \times \mathbf{S}_j$ terms in the low-energy SHG operators. Quantitative reproduction of the RA-SHG patterns and the spiral tilting angle suggests ground-state magnetoelectric effects likely do not dominate the SHG in NiI$_2$. 

Looking forward, the many-body cluster approach described in this letter may be easily extended to treat other non-linear optical probes and include any sufficiently local spin/orbital degrees of freedom. The strong frequency- and polarization-dependence of the dynamical operators allows, in principle, experiments to be designed \emph{a-priori} to probe specific correlation functions. Continued development of material-specific \emph{ab-initio} tools and microscopic understanding of excited-state processes and are key to realising this potential.

\subsection{Acknowledgments}
This material is based upon work supported by the National Science Foundation under Grant Number DMR-2338704. Calculations were performed using the Wake Forest University (WFU) High Performance Computing Facility. The authors acknowledge insightful discussions with R.~Valent\'i, C.~Kim, A. Salam and L.~Zhao.

\bibliographystyle{apsrev4-2}
\bibliography{nonlinear}

\clearpage

\section{End Matter}

\subsection{Cluster Calculations}
We consider an orbital basis composed of $d$-orbitals on the two Ni atoms supplemented by NTOs constructed as described below. The many-body Hamiltonian is:
\begin{align}
    \mathcal{H}_0 = \mathcal{H}_{\rm 1p} + \mathcal{H}_{\rm 2p} + \mathcal{H}_{\rm DC}
\end{align}
where $\mathcal{H}_{\rm 1p}$ is the one-particle Hamiltonian containing all hopping, crystal field, and spin-orbit coupling contributions, $\mathcal{H}_{\rm 2p}$ is the two-particle Hamiltonian containing Coulomb interactions between the Ni $d$-orbitals, and $\mathcal{H}_{\rm DC}$ is a one-particle double-counting correction applied to the metal orbitals. The one-particle Hamiltonian is:
\begin{align}
    \mathcal{H}_{\rm 1p} = \sum_{ij\alpha\sigma\beta\sigma^\prime} t_{ij}^{\alpha\sigma\beta\sigma^\prime} c_{i,\alpha,\sigma}^\dagger c_{j,\beta,\sigma^\prime}
\end{align}
where $c_{i,\alpha,\sigma}^\dagger$ creates an electron at atomic site $i$, centered at position $\mathbf{r}_i$, with orbital and spin quantum numbers $\alpha$ and $\sigma$, and $t_{ij}^{\alpha\sigma\beta\sigma^\prime}$ are hopping integrals. There are various gauge choices for the light-matter operators \cite{woolley2020power,schuler2021gauge}. For convenience, we employ a form with discrete translational invariance where $\hat{\vec{\mathcal{L}}}(\omega)  \equiv  \hat{\vec d}+\frac{i}{\hbar\omega}\hat{\vec j}$ is the sum of site-centered dipole and current terms, respectively:
\begin{align}
\hat{\vec d} = & \ -e\sum_{ij\alpha\sigma\beta\sigma^\prime} \langle w_{i,\alpha,\sigma}| (\hat{\mathbf{r}}-\mathbf{r}_j)|w_{j,\beta,\sigma^\prime}\rangle c_{i,\alpha,\sigma}^\dagger c_{j,\beta,\sigma^\prime} & \\
 \hat{\vec j} = & \ ie \sum_{ij\alpha\sigma\beta\sigma^\prime} t_{ij}^{\alpha\sigma\beta\sigma^\prime}(\mathbf{r}_i-\mathbf{r}_j) \ c_{i,\alpha,\sigma}^\dagger c_{j,\beta,\sigma^\prime} 
\end{align}
These may be computed from matrix elements of position operators and hopping integrals obtained from Wannier fitting. While the site-centered dipole operators are formally required to maintain gauge invariance at lowest order, we find they make negligible contribution to the SHG response at optical frequencies for NiI$_2$, and therefore have omitted them in all subsequent calculations.

To obtain the one-particle terms, we first performed band structure calculations at the GGA+SOC level using Fleur \cite{fleurCode,fleurWeb} and the structure reported in \cite{ketelaar1934kristallstruktur}. For this purpose, we employ a $7\times 7 \times 7$ $k$-point mesh for BZ sampling, with an augmented plane wave basis with $K_{\rm max} = 3.9 \ a_B^{-1}$, $G_{\rm max} = 11.7 \ a_B^{-1}$, and an angular momentum cutoff $\ell_{\rm max} = 10$. We then employed wannier90 \cite{freimuth2008maximally,mostofi2014updated} to obtain Wannier functions based on a $4 \times 4 \times 4$ $k$-point grid / real-space supercell. We construct two different sets of Wannier functions. In the first case, we employ both Ni $d$- and I $p$-orbital projections to fit all valence bands. We refer to this as the full $(d+p)$ basis. In the second case, we employ only Ni $d$-orbital projections to fit the bands in the vicinity of the Fermi level. We refer to this as the $(d)$-basis. For the projection axes defining the orientation of the orbitals, we choose the approximate cubic axes (denoted by $(\bar{x},\bar{y},\bar{z})$ to distinguish them from the ($x,y,z$) and ($X,Y,Z$) coordinates) such that $\bar{x}+\bar{y}+\bar{z}$ is parallel to the crystallographic $c$-axis, and the angle between the N-I bond directions and cubic axes is otherwise minimized. In each case, we employ zero localization steps to prevent spurious symmetry breaking. The resulting Wannier functions $w_{n\mathbf{R}}(\mathbf{r})$ are defined by unitary transformations $U_{mn}^\mathbf{k}$ on the Bloch functions $\psi_{m\mathbf{k}}(\mathbf{r})$:
\begin{align}
    w_{n\mathbf{R}}^{(d+p)}(\mathbf{r}) = & \ \sum_{m\mathbf{k}} U_{n m\mathbf{k}}^{(d+p)} \psi_{m\mathbf{k}}(\mathbf{r}) e^{-i\mathbf{k}\cdot\mathbf{R}}
    \\
        w_{n\mathbf{R}}^{(d)}(\mathbf{r}) = & \ \sum_{m\mathbf{k}} U_{n m\mathbf{k}}^{(d)} \psi_{m\mathbf{k}}(\mathbf{r}) e^{-i\mathbf{k}\cdot\mathbf{R}}
\end{align}
where $\mathbf{R}$ denotes a lattice vector, $n$ enumerates the spin/orbital quantum numbers of the Wannier functions, and $m$ denotes the index for bands included in the fitting. The dimensions of these Wannier bases differ; for $N$ momentum points, the dimension of the $(d)$-basis is $N D_{(d)}$, and the dimension of the $(d+p)$ basis is $N D_{(d+p)}$. We therefore supplement the $(d)$-basis by a set of auxilliary functions $w^{\rm (aux)}_{\alpha^\prime \mathbf{R}^\prime}$ obtained by appending an appropriate number of random vectors to $w_{n^\prime\mathbf{R}}^{(d)}$ and performing a Gram–Schmidt orthonormalization. This provides a unitary transformation:
\begin{align}
        \mathcal{U}_{n^\prime n}^{\mathbf{R}^\prime,\mathbf{R}} = \left\{ \begin{array}{cc}\left\langle w_{n^\prime\mathbf{R^\prime}}^{(d)}(\mathbf{r}) \right|\left. w_{n\mathbf{R}}^{(d+p)}(\mathbf{r})\right\rangle& n^\prime \leq D_{(d)}
        \\
        \left\langle 
        w^{\rm (aux)}_{n^\prime\mathbf{R^\prime}}(\mathbf{r}) | w_{n\mathbf{R}}^{(d+p)}(\mathbf{r})\right\rangle& n^\prime > D_{(d)}
        \end{array}
        \right.
\end{align}
The auxiliary $w^{\rm (aux)}_{n^\prime \mathbf{R}^\prime}$ functions will ultimately serve as a basis for the ligand-centered natural transition orbitals (NTOs), so their random definition is of no consequence.

The NTOs are obtained starting from the transition density matrices \cite{martin2003natural}:
\begin{align}
    \left[\mathbb{T}\right]_{i\alpha\sigma,j\beta\sigma^\prime} = & \ \langle \Psi_{\rm ex}| c_{i\alpha,\sigma}^\dagger c_{j\beta,\sigma^\prime}|\Psi_{\rm gs}\rangle
\end{align}
where $|\Psi_{\rm gs}\rangle$ and $|\Psi_{\rm ex}\rangle$ are ground and excited states, respectively. For a polarization $\mu$, taking $|\Psi_{\rm ex}\rangle = \frac{\hbar\omega}{e}\hat{\mathcal{L}}_\mu(\omega)|\Psi_{\rm gs}\rangle$ and restricting the light-matter operator to contain only current terms yields:
\begin{align}
    \left[\mathbb{T}_\mu\right]_{i\alpha\sigma,j\beta\sigma^\prime}
    = & \ t_{ji}^{\beta\sigma^\prime \alpha\sigma}(r^\mu_j-r^\mu_i) \langle \Psi_{\rm gs}| (1-\hat{n}_{i\alpha\sigma})\hat{n}_{j\beta\sigma^\prime} |\Psi_{\rm gs}\rangle
\end{align}
To define NTOs that span the \emph{whole} LMCT space for a cluster, we have $\alpha,\sigma$ include the 20 $d$-spin-orbitals on the Ni sites within the cluster (which we take to be empty $\langle \hat{n}_{i\alpha}\rangle = 0$), and have $\beta$ include all 768 I $p$ spin-orbitals in the entire $4\times 4\times 4$ supercell on which the Wannier functions are defined. We then perform the SVD on the transition density matrix which yields:
\begin{align}
    \mathbb{T}_\mu = \mathbb{U}_\mu \Lambda_\mu \mathbb{V}_\mu^\dagger
\end{align}
where $\Lambda_\mu$ is a $20 \times 768$ rectangular matrix containing the singular values on the main diagonal, and $\mathbb{V}_\mu^\dagger$ defines a unitary transformation on the $p$ spin-orbitals corresponding to the SVD. For NiI$_2$, we take the first 20 vectors in each of $\mathbb{V}_x^\dagger$ and $\mathbb{V}_y^\dagger$, in order to capture the LMCT excitations accessible via in-plane polarized light according to the normal incidence geometry of the experiments in \cite{song2022evidence}. Since the vectors for different polarizations are not required to be orthonormal, we perform symmetric orthonormalization to yield 40 spin-NTOs. All single-particle operators $\mathcal{H}_{\rm 1p}$ and $\mathcal{H}_{\rm lm}$ are rotated into this ($d$+NTO) basis.

For the Ni Coulomb terms, we use:
\begin{align}
\mathcal{H}_{\rm 2p} = \sum_{i\alpha\beta\delta\gamma}\sum_{\sigma\sigma^\prime}U_{\alpha\beta\gamma\delta} \ c_{i,\alpha,\sigma}^\dagger c_{i,\beta,\sigma^\prime}^\dagger c_{i,\gamma,\sigma^\prime} c_{i,\delta,\sigma}
\end{align}
where $U_{\alpha\beta\gamma\delta}$ were parameterized by the Slater parameters $F_0^{dd}, F_2^{dd}, F_4^{dd}$ \cite{sugano1970multiplets}. These may be related to the more familiar Hubbard $U_{t2g}$ and Hunds $J_{t2g}$ Kanamori parameters for the $t_{2g}$ orbitals via $U_{t2g} =  F_0^{dd} + \frac{4}{49}\left(F_2^{dd}+F_4^{dd}\right)$ and $J_{t2g} = \frac{1}{49}\left(3 F_2^{dd}+5F_4^{dd}\right)$. We constrain $F_4^{dd}/F_2^{dd} = 5/8$ (which is respected approximately by $3d$ elements), and employ $U_{t2g} = 4.0$ eV, $J_{t2g} = 0.8$ eV. This corresponds to $F_0^{dd} = 2.82$ eV, $F_2^{dd} = 8.93$ eV, and $F_4^{dd} = 5.58$ eV. The latter two parameters are reduced to $\sim 73\%$ from their atomic Hartree-Fock values \cite{haverkort2005spin}, which accounts for both screening and ligand-metal hybridization. The parameters are similar to those derived from spectroscopic studies \cite{zaanen1986determination,occhialini2024nature}. For the double-counting correction, we employ the ``Around Mean-Field'' scheme:
\begin{align}
    \mathcal{H}_{\rm DC} = 
\sum_{i\alpha\beta\gamma}\sum_{\sigma}\left( 2 U_{\alpha\gamma\gamma\beta} - U_{\alpha\gamma\beta\gamma}\right)  \langle n_d\rangle \  c_{i,\alpha,\sigma}^\dagger c_{i,\beta,\sigma} 
\end{align}
where $\langle n_d \rangle = 4/5$ is the average occupancy of a $d$ spin-orbital in $d^8$ filling. The results for the experimental frequency $\hbar\omega = 1.25$ eV are not strongly sensitive to the choice of Coulomb parameters because neither $\omega$ nor $2\omega$ is resonant with the intersite $d$-$d$ excitations.

We perform an exact diagonalization of the many-body Hamiltonian $\mathcal{H}_0$ in the ($d$+NTO) Fock space. We obtain the lowest $(2S+1)^2 = 9$ many-body states $|\Psi_n\rangle$ of the two-site cluster, nominally corresponding to the combinations of $m_S = +1,0,-1$ states on the two Ni sites with $t_{2g}^6 e_g^2$ filling. To identify the low energy states $|\Psi_n\rangle$ with specific $m_S$ states, we follow the des Cloizeaux effective Hamiltonian procedure \cite{des1960extension}. We project the low-energy Fock states onto pure spin states, and perform a symmetric orthonormalization to define projected states $|\Phi_n\rangle$:
\begin{align}
    |\Phi_n\rangle = & \  \mathbb{S}^{-1/2} \ \mathbb{P}|\Psi_n\rangle
    \\
    \mathbb{S} = & \ \sum_{n,m}|\Psi_n\rangle\langle \Psi_n| \mathbb{P}|\Psi_m\rangle\langle\Psi_m|
    \\
    \{|m_S^i,m_S^j\rangle\} = & \ \mathbb{U} \{ |\Phi_n\rangle\}
\end{align}
where $\mathbb{P} = \sum_{m_S^i,m_S^j} |m_S^i,m_S^j\rangle\langle m_S^i,m_S^j |$ is the pure spin projector and $\mathbb{S}$ is the overlap matrix. The projected states $\{|\Phi_n\rangle\}$ are orthonormal and span precisely the same space as the pure spin states, so that $\mathbb{U} \ \mathbb{S}^{-1/2} \ \mathbb{P}$ defines a unitary mapping between the true many-body low-energy Fock space and the pure spin states. Approximately speaking, this mapping assigns a label of $|m_S^i,m_S^j\rangle$ to the linear combination of $\{|\Psi_n\rangle\}$ low-energy states that has maximum overlap with $|m_S^i,m_S^j\rangle$. As a consequence, any operator in the low-energy space can be mapped to pure spin operators. 

To obtain the dynamical SHG operators as spin-operators, we use the band Lanczos approach \cite{weikert1996block,meyer1989band} to construct a Krylov spaces for $\hat{\mathcal{L}}_\mu \{|\Psi_n\rangle\}$, and use these spaces to evaluate eq'n (3) of the main text as a function of $\omega$, employing a fixed broadening of the poles. 
This yields the matrix elements $\langle \Psi_n | \hat{\mathcal{O}}_{\mu\xi\nu} (\omega) |\Psi_m\rangle$ between low-energy Fock states. We then apply the mapping to yield spin operators at each computed frequency.

\subsection{SHG Fitting Functions}
We define:
\begin{align}
    A_{\mu\xi\nu}^\gamma = \left[\mathbf{C}_{ij}^{\mu\xi\nu}(\omega) + \left(\mathbf{C}_{ij}^{\mu\xi\nu} (-\omega)\right)^\dagger\right]_\gamma
\end{align}
as the cartesian $\gamma$-component of $\mathbf{C}_{ij}(\omega) + \mathbf{C}_{ij}^\dagger (-\omega)$ for the nearest neighbor bond depicted in Fig.~\ref{fig:orb} of the main text evaluated at $\hbar\omega = 1.25$ eV, for incoming polarizations $\xi,\nu$ and outgoing polarization $\mu$. For a single incommensurate spiral domain, in the parallel configuration $\mu = \xi = \nu$, the SHG susceptibility is evaluated as:
    \begin{align}
   \chi_{||}(\theta) \propto & \ 
  \left\{ \left[\left(
   A_{xxy}^y + A_{xyx}^y + A_{yxx}^y - A_{yyy}^y
   \right)\left(4\cos 2\theta  +1\right)\right. \right.
      \nonumber \\
      & \ \left. +3 A_{xxx}^x + A_{xyy}^x + A_{yxy}^x + A_{yyx}^x - 4A_{yyy}^y
   \right]\sin\alpha 
   \nonumber \\
   & \ \left.-2 \left(
   3 A_{xxx}^z + A_{xxy}^z + A_{yxy}^z + A_{yyx}^z
   \right)\cos \alpha
   \right\}  \sin\theta
\end{align}
where $\theta$ is the angle between the in-plane propagation vector $\mathbf{Q}$, and $\alpha$ is the angle between the normal vector of the spiral plane and the crystallographic $c$-axis [see Fig.~\ref{fig:structure}(b)]. Similarly, in the perpendicular configuration $\mu \perp \xi = \nu$, we have:
\begin{align}
    \chi_\perp(\theta) \propto & \ \left\{\left[ \left(A_{xxy}^y + A_{xyx}^y + A_{yxx}^y - A_{yyy}^y\right)(4\cos 2\theta-1)\right.\right.
    \nonumber
    \\
    & \ \left. 
    3 A_{xyy}^x+ A_{xxx}^x - A_{yxy}^x - A_{yyx}^x 
    -  4 A_{yxx}^y 
    \right]\sin\alpha
    \nonumber \\
    & \ \left. 
    -2 (3 A_{xyy}^z +A_{xxx}^z - A_{yxy}^z - A_{yyx}^z  )\cos\alpha
    \right\} \cos\theta
\end{align}
Anisotropy of the experimental RA-SHG data from \cite{song2022evidence} indicates a mixture of multiple domains with propagation vectors related by three-fold rotation. We therefore fit the data to:
\begin{align}
I \propto |\chi(\theta)+d_2\chi(\theta+2\pi/3)+d_3\chi(\theta+4\pi/3)|^2
\end{align}
with $d_2$ and $d_3$ giving the relative amount of additional domains.

\end{document}